\documentclass[journal]{IEEEtran}
\ifCLASSINFOpdf

\else

\fi
\usepackage[cmex10]{amsmath}
\usepackage{amssymb}
\usepackage{amsfonts}
\usepackage{cite}
\usepackage{graphicx}
\usepackage{array,color}
\usepackage{multirow}
\usepackage{amsmath}
\usepackage{stfloats}
\usepackage{graphicx}
\usepackage{subfigure}
\usepackage{tabularx}
\usepackage{epsfig,epsf,color,balance,cite}
\usepackage{setspace}
\usepackage{bm}
\usepackage{algorithm}
\usepackage{algorithmic}
\usepackage{textcomp}
\usepackage{subfigure}
\usepackage{caption}
\usepackage{graphicx}
\usepackage{setspace}
\begin{document}
\title{Long-Term CSI-based Design for RIS-Aided Multiuser MISO Systems Exploiting Deep Reinforcement Learning}
\author{ Hong Ren, Cunhua Pan,  Liang Wang, Zhoubing Kou, and Kezhi Wang
\thanks{(Corresponding author: Cunhua Pan) H. Ren and Zhoubing Kou are with   Southeast University, Nanjing 210096, China. (e-mail: {hren, zbkou}@seu.edu.cn).  C. Pan is with Queen
Mary University of London, London E1 4NS, U.K. (e-mail: c.pan@qmul.ac.uk).  K. Wang and Liang Wang are with Northumbria University, UK. (e-mail:
{kezhi.wang, Liang.wang}@northumbria.ac.uk).}
%\thanks{This work was supported by ...}
}

\maketitle
\begin{abstract}
In this paper, we study the transmission design for reconfigurable intelligent surface (RIS)-aided multiuser communication networks. Different from most of the existing contributions, we consider  long-term CSI-based transmission design, where both the beamforming vectors at the base station (BS) and the phase shifts at the RIS are designed based on long-term  CSI, which can significantly reduce the channel estimation overhead. Due to the lack of explicit ergodic  data rate expression, we propose a novel deep deterministic policy gradient (DDPG) based algorithm to solve the optimization problem, which was trained by using the channel vectors generated in an offline manner. Simulation results demonstrate that the achievable net throughput is higher than that achieved by the conventional instantaneous-CSI based scheme when taking the channel estimation overhead into account.
\end{abstract}

\IEEEpeerreviewmaketitle
%\newpage
\section{Introduction}

Due to its appealing features of low hardware cost and energy consumption, reconfigurable intelligent surface (RIS) has been widely regarded as one of the promising technologies in the future wireless networks \cite{qingqing,Chongwentwc,pan-mag}.  To reap the benefits brought by the RIS, the active beamforming at the base station (BS) and  phase shifts at the RIS should be carefully designed. Hence, extensive research efforts have been devoted to the transmission design for RIS-aided various wireless communication systems, such as multicell networks \cite{cunhuatwc},  simultaneous wireless information and power transfer (SWIPT) \cite{cunhuajsac}, physical layer security \cite{hongsheng},  and mobile edge computing \cite{tongbai}. The above contributions were based on the assumption that perfect instantaneous channel state information (CSI) is available at the BS, which will entail excessive amount of channel estimation overhead that is  proportional to the number of reflecting elements.

One alternative method to address the above challenge is to adopt the long-term CSI \cite{yuhan,mingminzhao,kangdazhi,Papazafeiropoulos,kangdacl,hu2020statistical},  which varies much more slowly than the instantaneous CSI and is relatively easy to obtain.  Specifically, the contributions in \cite{yuhan,mingminzhao,kangdazhi,Papazafeiropoulos,kangdacl} focused on the two-timescale design, in which the phase shifts at the RIS were designed based on the long-term CSI such as angle information or/and channel covariance matrices, while the BS still needs  the instantaneous aggregate CSI for the active beamforming design.  Most recently, the authors in \cite{hu2020statistical} proposed a transceiver design, where both the active beamforming at the BS and phase shifts were designed based on the long-term CSI. However, only one user was considered in \cite{hu2020statistical}, and closed-form solution can be readily derived.  Most recently, the authors studied the ergodic sum rate maximization problem in \cite{jinghe2021,ganxu2021} for multi-user communication systems based on fully long-term CSI and Jensen's inequality was adopted to approximate the ergodic data rate expression. However, the fairness issue cannot be guaranteed in \cite{jinghe2021,ganxu2021}.

Against the above background, the main contributions of this paper are summarized as follows:
\begin{itemize}
  \item  We first consider the fairness design for RIS-aided multi-user communications based on long-term CSI with the aim of maximizing the minimum ergodic user data rate.
  \item Unlike the single-user case in \cite{hu2020statistical},  the explicit closed-form expression of the ergodic data rate cannot be obtained. In addition, the max-min objective function is non-differentiable.    To address this issue, we propose a novel deep deterministic policy gradient (DDPG) based algorithm to solve the optimization problem. In specific, we first estimate the long-term  CSI. Then, we generate a set of  channels based on the Rician distribution, which were used to train the DDPG networks. The solution produced by the DDPG networks can then be used in the remaining channel coherence time intervals where the long-term CSI keeps fixed.
\end{itemize}

\section{System Model}\label{system}
Consider a downlink MISO system where an $M$-antenna BS  communicates with $K$ single-antenna users. An RIS with $N$ reflecting elements  is deployed between the users and the BS to improve the communication quality. The RIS configuration matrix can be denoted as $\mathbf{\Phi}=\text{diag}\left\{\phi_1,\phi_2,\cdots,\phi_N\right\}$, where $\phi_n$ is the phase shift of the $n$-th element and $|\phi_n|=1, n=1,2,\cdots,N$. Let us denote $\mathbf{h}_k$ and $\mathbf{g}_k$ as the channel vector between the $k$-th user and the RIS  and  the direct channel between the $k$-th user and the BS, respectively. The channel matrix between the BS and the RIS is denoted as $\mathbf{G}$.

%\vspace{-0.3cm}
%\begin{figure}[h]
%	\centering
%	\includegraphics[width=0.3\textwidth]{fig/systemmodel.pdf}
%	\caption{A RIS-aided Downlink MISO system}
%	\label{system}
%\end{figure}
%\vspace{-0.3cm}

The beamforming vector for the $k$-th user is denoted as $\mathbf{w}_k$. The set of all beamforming vectors is denoted as $\mathbf{W}=\left[ {{{\bf{w}}_1}, \cdots ,{{\bf{w}}_K}} \right]$. Then, the transmitted complex baseband signal $\mathbf{x}\in\mathbb{C}^{M\times 1}$ is given by $\mathbf{x}=\mathbf{W}\mathbf{s}$, where  $\mathbf{s}\in \mathbb{C}^{K\times 1}=[s_1,s_2,\dots,s_K]^T$ denotes data streams to $K$ users, and ${\mathbb{E}}\{{\bf{s}\bf{s}}^H\}={\bf{I}}_K$.

 The received signal of the $k$-th user is given by
\begin{equation}
y_k=({\mathbf{h}_k^T\mathbf{\Phi} \mathbf{G}}+{\mathbf{g}_k^T})\sum\nolimits_{j=1}^{K}\mathbf{w}_js_j+n_k,
\end{equation}
where   $n_k\sim\mathcal{CN}(0,\sigma^2)$ is the additive white Gaussian noise (AWGN) at the $k$-th user.

The Rician channel model is adopted, and thus all the channels can be represented as
\begin{align}
\mathbf{G}&=\sqrt{\kappa }{\left(\sqrt{\frac{\delta}{\delta+1}}\bar{\mathbf{G}}+\sqrt{\frac{1}{\delta+1}}\tilde{\mathbf{G}}\right)},\\
\mathbf{g}_k&=\sqrt{\beta_k}\left(\sqrt{\frac{\varepsilon_k}{\varepsilon_k+1}}\bar{\mathbf{g}}_k+\sqrt{\frac{1}{\varepsilon_k+1}}\tilde{\mathbf{g}}_k\right),\\
\mathbf{h}_k&=\sqrt{\gamma_k}\left(\sqrt{\frac{\eta_k}{\eta_k+1}}\bar{\mathbf{h}}_k+\sqrt{\frac{1}{\eta_k+1}}\tilde{\mathbf{h}}_k\right),
\end{align}
where $\kappa$ denotes the large-scale path-loss factor  from the BS to the RIS, $\beta_k$ and $\gamma_k$ are the large-scale path-loss factors from the RIS and the BS to the $k$-th user, respectively. $\tilde{\mathbf{G}}$, $\tilde{\mathbf{g}}_k$ and $\tilde{\mathbf{h}}_k$ denote the non-LoS components that represent the scattering components, each element of which follows  zero mean and unit variance complex Gaussian distribution. $\delta$, $\varepsilon_k$ and $\eta_k$ are Rician factors. $\bar{\mathbf{G}}$, $\bar{\mathbf{g}}_k$ and $\bar{\mathbf{h}}_k$ represent LoS components that are given by uniform linear antenna array response as follows
\begin{equation}
\overline {\bf{G}}  = \sum\limits_{i = 1}^I {{{\bf{a}}_N}({\phi _{AoA,i}}){\bf{a}}_M^T({\phi _{AoD,i}})}, 
\end{equation}
\begin{equation}
\bar{\mathbf{g}}_k=\mathbf{a}_N(\theta_{AoD}^k),
\end{equation}
\begin{equation}
\bar{\mathbf{h}}_k=\mathbf{a}_M(\psi_{AoD}^k),
\end{equation}
where $I$ is the number of links from the BS to the RIS, $\phi_{AoA}$ denotes the arrival of angle (AoA) from the BS to the RIS, $\phi_{AoD}$ is the angle of departure (AoD) from the BS to the RIS, $\theta_{AoD}^k$ is the AoD from the RIS to the $k$-th user, $\psi_{AoD}^k$ is the AoD from the BS to the $k$-th user, and $\mathbf{a}_x(\theta)=[1,e^{j\theta},\dots,e^{j(x-1)\theta}]^T$. Therefore, $\bar{\mathbf{G}}$, $\bar{\mathbf{g}}_k$ and $\bar{\mathbf{h}}_k$ are only related to the AoD and AoA of signals that are named as long-term CSI.

Herein, the $k$-th user's instantaneous signal-to-interference-plus-noise ratio (SINR) can be written as
\begin{equation}
{\text{SINR}}_{k}=\frac{|({\mathbf{h}_k^T\mathbf{\Phi} \mathbf{G}}+{\mathbf{g}_k^T})\mathbf{w}_{k}|^2}{\sum_{j=1,j\ne k}^{K}|({\mathbf{h}_k^T\mathbf{\Phi} \mathbf{G}}+{\mathbf{g}_k^T})\mathbf{w}_{j}|^2+\sigma_k^2}.
\label{equation2_10}
\end{equation}
The instantaneous data rate of the $k$-th user is given by
\begin{equation}\label{jerofh}
	R_k=\log_2(1+{\text{SINR}}_k).
\end{equation}

In this paper, we aim to jointly optimize the precoding matrix $\mathbf{W}$ and the configuration matrix $\mathbf{\Phi}$ to maximize the minimum ergodic transmission data rate of $K$ users, where the expectation is taken over the  non-LoS components of all channels. Mathematically, the optimization problem is formulated as
\begin{equation}
\begin{aligned}
&\underset {\mathbf{W},\mathbf{\Phi}}{\max}\;\underset{k}{\min}\;\;\;\mathbb{E}[R_k]\\
&\;\;\;\;\;\;\text{s.t.} \ \quad\;\;\text{C1}: {\rm{tr}}\{\mathbf{W}\mathbf{W}^H\}\le P_t,\\
&\;\;\;\;\;\;\;\;\;\quad\quad\;\text{C2}:  |\phi_n|=1,\forall n=1,2,\dots,N,\\
\end{aligned}
\label{opt-model}
\end{equation}
where C1 is the power constraint at the BS and $P_t$ is the maximum power limit, C2 represents unit-modulus constraint of the reflecting elements.
Since the closed-form expression of the ergodic data rate is not available,    Problem (\ref{opt-model}) is challenging to solve. In addition, the max-min objective function is non-differentiable and the existing algorithms developed for sum rate maximization problem in \cite{jinghe2021,ganxu2021} are not applicable.

\section{Proposed Algorithm}\label{proformu}
In this section, we first introduce the transmission scheme and then provide a DDPG based algorithm to solve the problem.
\vspace{-0.2cm}
\subsection{Transmission Scheme}
\begin{figure}[h]
	\centering
	\includegraphics[width=0.5\textwidth]{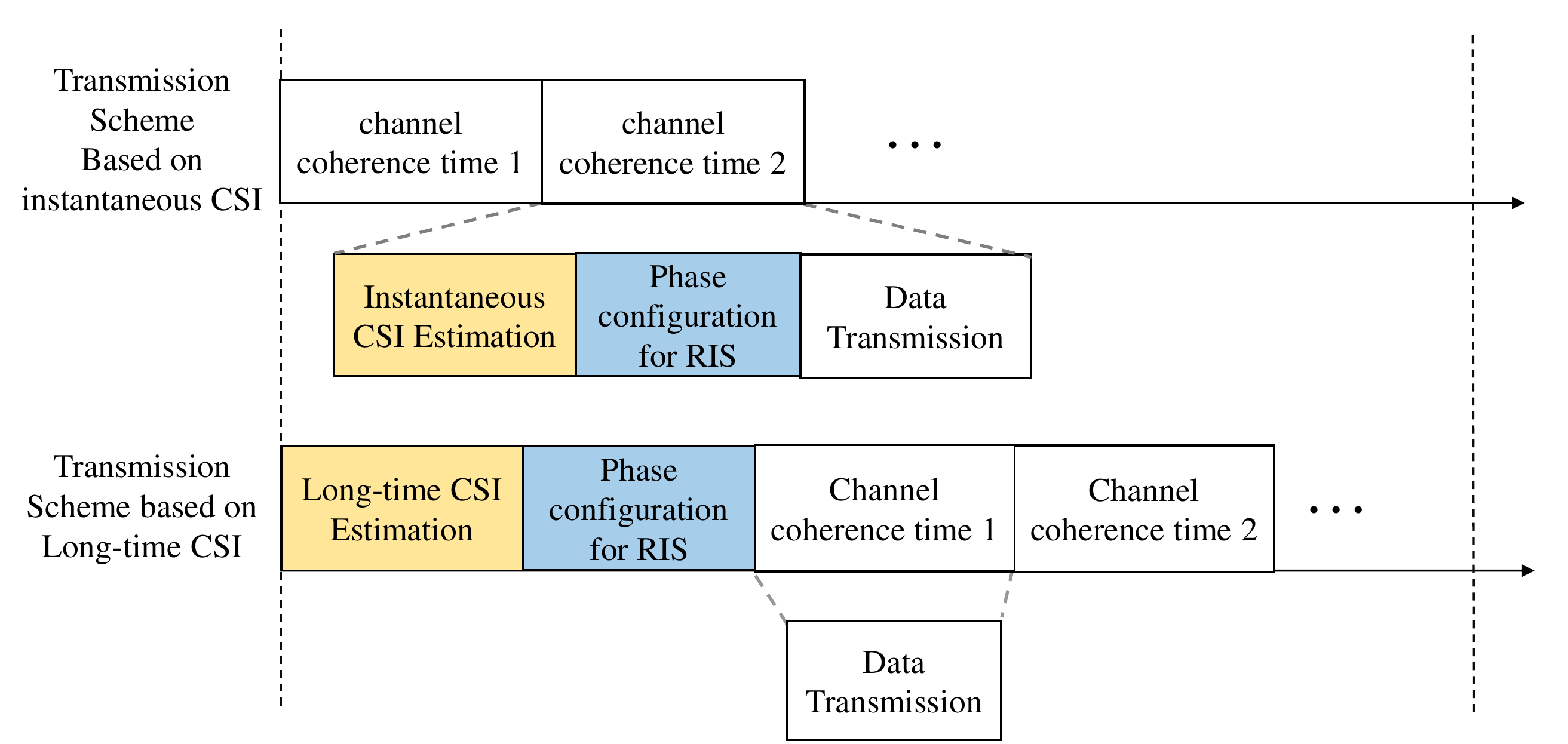}
	\caption{Two Transmission Schemes}
	\label{operationmecha}
\vspace{-0.5cm}
\end{figure}

In this paper, we consider a series of consecutive channel coherence time intervals (CCTIs) as shown in Fig. \ref{operationmecha}, in which the long-term CSI  keeps invariant and non-LoS components vary in each CCTI.

Most of the existing contributions considered the first transmission scheme as shown in Fig. \ref{operationmecha}, where the precoding matrix and the phase shift matrix are designed based on the instantaneous CSI. In specific, in each CCTI, the instantaneous CSI should be estimated, based on which the beamforming and phase shift should be optimized. Finally, the data is transmitted in the remaining part of the CCTI. However, this scheme has three drawbacks. First, the training overhead is high as it needs to estimate the instantaneous cascaded BS-RIS-user channel and the direct BS-user channel, the amount of which is proportional to the number of reflecting elements that is excessive. Second, in each CCTI, the BS needs to calculate the precoding matrix and phase shift matrix, which incurs high computational complexity. Third, in each CCTI, the calculated phase shift matrix needs to be sent back to the RIS controller, which induces high feedback overhead.

To address the above issues, this paper considers the long-term CSI-based transmission scheme   shown in the lower part of Fig. \ref{operationmecha}. In this scheme, the BS first estimates the long-term CSI (angle information). Based on the estimated angle information,  the precoding matrix and phase shift matrix are designed, which will be used in the remaining CCTIs. Note that the precoding matrix and the phase shift matrix  only need to be calculated  once in the considered time intervals. In addition, the long-term CSI is only estimated  once, which significantly reduces the channel estimation overhead. Note that in the considered time intervals, all the time slots are dedicated for data transmission.

Next, we derive the ergodic net data rate by considering the channel estimation overhead. Assume that there are a total of $T$ CCTIs, and there are $\tau_c$ time slots in each CCTI.

 For the instantaneous CSI-based scheme, one needs to estimate the cascaded channel matrix $\mathbf{G}_k=\mathbf{G}\rm{diag}\{\mathbf{h}_k\}$ and the direct channel matrix $\mathbf{g}_k$ for each CCTI. Based on \cite{liangliu2020}, the minimum number of time slots for channel training that is denoted as $\tau$ should satisfy:
\begin{equation}\label{ftrgrt}
  \tau \ge 2K+N-1.
\end{equation}
Then, at the $t$-th CCTI, the net instantaneous data rate is given by
\begin{equation}\label{rgsr}
 R_{k,{\rm{ins}}}^{(t)} = \left( {1 - \frac{{2K + N - 1}}{{{\tau _c}}}} \right){\log _2}\left( {1 + {\rm{SINR}}_{k,{\rm{ins}}}^{(t)}} \right),
\end{equation}
where ${\rm{SINR}}_{k,{\rm{ins}}}^{(t)}$ is given by
\[{\rm{SINR}}_{k,{\rm{ins}}}^{(t)} = \frac{{|({\bf{h}}_k^{(t)T}{{\bf{\Phi }}^{(t)}}{{\bf{G}}^{(t)}} + {\bf{g}}_k^{(t)T}){\bf{w}}_k^{(t)}{|^2}}}{{\sum\limits_{j = 1,j \ne k}^K | ({\bf{h}}_k^{(t)T}{{\bf{\Phi }}^{(t)}}{{\bf{G}}^{(t)}} + {\bf{g}}_k^{(t)T}){\bf{w}}_j^{(t)}{|^2} + \sigma _k^2}}.\]

For the long-term CSI based design, we ignore the channel estimation overhead incurred by the angle information estimation as it can be estimated with limited overhead and only needs to be estimated at the start of the considered time intervals. Hence, at the $t$-th CCTI, the net instantaneous data rate is given by
\begin{equation}\label{rsfre}
  R_{k,{\rm{sta}}}^{(t)} = {\log _2}\left( {1 + {\rm{SINR}}_{k,{\rm{lon}}}^{(t)}} \right).
\end{equation}
where ${\rm{SINR}}_{k,{\rm{lon}}}^{(t)}$ is given by
\[{\rm{SINR}}_{k,{\rm{lon}}}^{(t)} = \frac{{|({\bf{h}}_k^{(t)T}{\bf{\Phi }}{{\bf{G}}^{(t)}} + {\bf{g}}_k^{(t)T}){{\bf{w}}_k}{|^2}}}{{\sum\limits_{j = 1,j \ne k}^K | ({\bf{h}}_k^{(t)T}{\bf{\Phi }}{{\bf{G}}^{(t)}} + {\bf{g}}_k^{(t)T}){{\bf{w}}_j}{|^2} + \sigma _k^2}}.\]

\vspace{-0.5cm}
{\subsection{DRL description}}

As shown at the lower part of Fig.~\ref{operationmecha}, one algorithm needs to be developed to solve the optimization problem based on the long-term CSI. Since the closed-form analytical expression of the ergodic data rate is not available, we propose a novel DDPG-based method to solve Problem (\ref{opt-model}).

In a standard architecture of DRL, the optimization problem can be viewed as the Markov Decision process (MDP) problem, which means an agent is deployed to interact with the environment in discrete time steps (time slots). In each time step, the agent consistently observes the state $s^{(t)}$, executes the action $a^{(t)}$, and receives a reward $r^{(t)}$ from the environment. During the training process, the agent learns to find the optimal policy $\pi(\bm{a}|\bm{s})$ that maps the action set $\bm{a}$ and state set $\bm{s}$, where $a^{(t)}\in\bm{a}, s^{(t)} \in \bm{s}$, aiming at maximizing the accumulated reward $R^{(t)}=\sum_{t'=t}^{T}\gamma^{t'-t}r^{(t')}$, where $T$ is the overall number of time steps, and $\gamma \in [0, 1]$ is the discount factor. Furthermore, a DNN is implemented to approximately calculate the Q-value, which can be expressed as
\begin{equation}
	\begin{aligned}
		Q(s^{(t)}, a^{(t)}) = \mathbb{E}\big[R^{(t)}|s^{(t)}, a^{(t)}\big],
	\end{aligned}
\end{equation}
where $\mathbb{E}\big[\cdot \big]$ denotes the expectation operator.

\subsection{Proposed Algorithm based on DDPG}

\begin{figure}[h]
	\centering
	\includegraphics[width=0.4\textwidth]{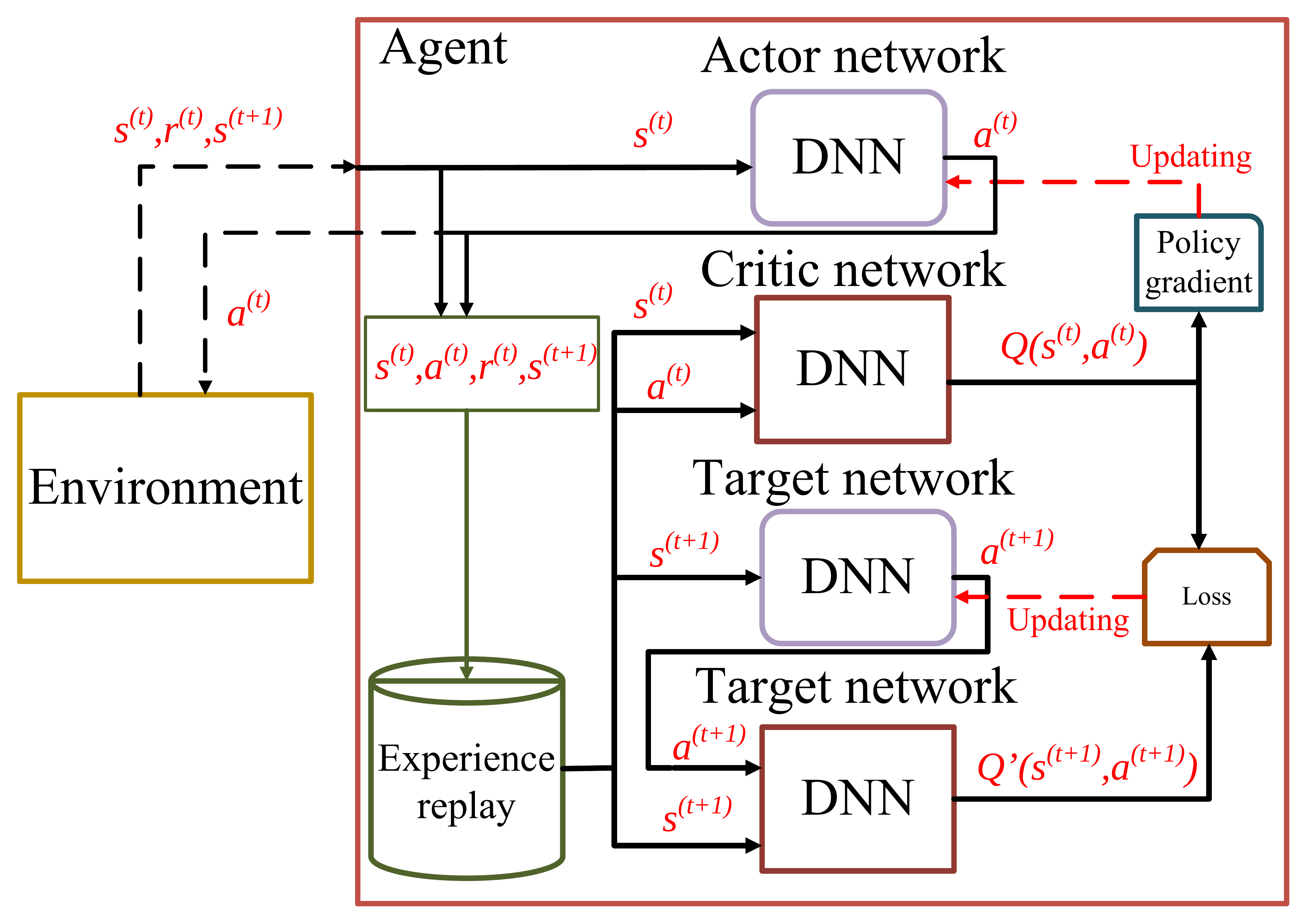}
	\caption{The DRL-based precoding matrix and phase shift design framework using DDPG}
	\label{ddpgfigure}
\end{figure}

In order to tackle the above optimization problem, we propose a novel algorithm based on DDPG for determining the phase shift of reflecting elements of RIS and the precoding matrix at the BS. The proposed framework is presented in Fig. \ref{ddpgfigure}.  To this end, we first generate a set of offline instantaneous CSI $\{\mathbf{G}^{(t)}, \mathbf{g}_k^{(t)}, \mathbf{h}_k^{(t)}, t=1,2,\cdots, \forall k\}$ based on the Rician fading distribution and the long-term CSI.

 Specifically, in the framework of our proposed algorithm, we view the system model as the environment, an agent is deployed at the BS. A DNN named actor $\pi(a^{(t)}, s^{(t)}|\omega^\pi)$ is deployed to generate the optimal action $a^{(t)}$ based on the  given state $s^{(t)}$, another DNN named critic $Q(s^{(t)}, a^{(t)}|\omega^Q)$ is deployed to generate Q-value, with the input of $a^{(t)}$ and $s^{(t)}$, where $\omega^\pi, \omega^Q$ are the parameters of actor and critic separately. Besides, both actor and critic have their target networks with the same structure, which can be expressed as $\pi'(a^{(t)}, s^{(t)}|\omega^{\pi'})$ and $Q'(s^{(t)}, a^{(t)}|\omega^{Q'})$ with parameters $\omega^{\pi'}$, $\omega^{Q'}$. For achieving better convergence performance and stabilizing the training process, a memory named experience replay is deployed to store the experience which consists of $\big[s^{(t)}, a^{(t)}, r^{(t)}, s^{(t+1)}\big]$. We define the state $s^{(t)}$, action $a^{(t)}$, and reward $r^{(t)}$ as follows:
\begin{enumerate}
	\item $\textbf{State}~s^{(t)}$: The state $s^{(t)}$ in the $t$-th time step is defined as a vector containing three parts. The first part is the set that includes the current real and imaginary part of precoding matrix $\bm{W}$; The second part is the set that contains the real part and imaginary part of the current phase shift of reflecting elements of RIS $\bm{\Phi}$; The third part is the set that has the real part and imaginary part of the CSI, i.e.,  $\mathbf{G}^{(t)}, \mathbf{g}_k^{(t)}, \mathbf{h}_k^{(t)}$, $k = 1,2,...,K$.
	
	\item $\textbf{Action}~{\bf{a}}^{(t)}$: The action ${\bf{a}}^{(t)}$ is defined as a vector in $\mathbb{R}^{2MK+N}$ that contains two kinds of components. Note that the activation function applied in the actor network is ${\rm{tanh}}(\cdot)$, the output value should be  within $[-1,1]$. Thus, the actor network can directly generate the real part and imaginary part of each element in precoding matrix $\bm{W}$. Then, we normalize and reformulate the optimized precoding matrix as follows:
	\begin{equation}
	\begin{aligned}
		\bm{W} \leftarrow \frac{\sqrt{P_t}\bm{W}}{||\bm{W}||_F}.
	\end{aligned}
	\end{equation}
	Additionally, as the continuous phase shift is applied in this paper and according to the Euler's formulation, the $n$-th phase shift $\phi_n$  can be represented by an angle $\phi^a_n$ within $[0, 2\pi]$, which can be expressed as
	\begin{equation}
	\begin{aligned}
		\phi_n = {\rm{cos}}(\phi^a_n)+j\cdot {\rm{sin}}(\phi^a_n).
	\end{aligned}
	\end{equation}
Then,  given the  $i$-th action value  in ${\bf{a}}^{(t)}$ (denoted as $a^{(t)}_i \in [-1, 1]$) and the current phase shift angle $\phi^a_n$, the optimized phase shift angle can be given by
\begin{equation}
\begin{aligned}
	\phi^a_n \leftarrow \phi^a_n + a^{(t)}_i \pi.
\end{aligned}
\end{equation}
For the above defined action vector, the constraints in the optimization problem  such as C1 and C2 are always satisfied.

	\item  $\textbf{Reward}~r^{(t)}$: The reward function is defined as follows:
	\begin{equation}\label{reward}
	\begin{aligned}
		{r^{(t)}} = \mathop {\min }\limits_k   {R_k^{(t)}},
	\end{aligned}
	\end{equation}
	where $R_k^{(t)}$ can be calculated by using Eq.~(\ref{jerofh}) and the $t$-th offline instantaneous CSI.
\end{enumerate}

We also give the overall algorithm design in Algorithm.~\ref{algo}.

\begin{algorithm}[!htpb]
	\caption{Proposed Algorithm}\label{algo}
	\begin{algorithmic}[1]
		\STATE Initialize actor $\pi(\cdot|\omega^\pi)$, critic $Q(\cdot|\omega^Q)$ networks with parameters $\omega^\pi$ and $\omega^Q$;\
		\STATE Initialize target networks with parameters $\omega^{\pi'} = \omega^\pi$ and $\omega^{Q'} = \omega^Q$;\
		\STATE Initialize experience replay memory with size $Z$, sample size $V$;\
		\FOR{epoch = 1,2,... }
		\STATE Initialize $s^{(t)}$;\
		\FOR{t = 1,2,...}
		\STATE Obtain $s^{(t)}$;\
		\STATE Obtain action $\pi(a^{(t)}, s^{(t)}|\omega^\pi)+\lambda \rho^a$;\
		\STATE Execute action $a^{(t)}$;\
		\STATE Obtain reward $r^{(t)}$ according to Eq.~(\ref{reward});\
		\STATE Store experience $\big[s^{(t)}, a^{(t)}, r^{(t)}, s^{(t+1)}\big]$ into experience replay memory;\
		\IF{the learning process starts}
		\STATE Randomly sample $V$ experiences from experience replay memory;\
		\STATE Update critic according to Eq.~(\ref{loss_fun});\
		\STATE Update actor according to Eq.~(\ref{policy_gra});\
		\STATE Update target networks with rate $\eta$:\
		\STATE $\omega^{\pi'} \leftarrow \eta \omega^{\pi} + (1-\eta) \omega^{\pi'} $ \
		\STATE $\omega^{Q'} \leftarrow \eta \omega^{Q} + (1-\eta) \omega^{Q'} $ \
		\ENDIF
		\ENDFOR
		\ENDFOR
	\end{algorithmic}
\end{algorithm}

Precisely, we first initialize the actor, critic networks, two target networks, and experience replay memory separately. In each training epoch, the state $s^{(t)}$ can be obtained from the environment. In Line 8, the action is generated by the actor network. Note that a random noise $\rho^a$ is added, which decays with rate $\lambda$, for better exploration. After executing the action, the reward $r^{(t)}$ can be obtained by Eq.~(\ref{reward}). Then, the experience is stored into the memory. When the learning process starts, $V$ experiences are randomly sampled for training the networks. The critic network is updated through minimizing the loss function, which is expressed as:
\begin{equation}\label{loss_fun}
\begin{aligned}
	\!\!\!\!L \!\!=\!\! \frac{1}{V} \!\!\sum_{i=1}^{V} \big(r^{i} \!\!+\!\gamma Q'(s^{(i+1)}, a^{(i+1)}|\omega^{Q'}\!)\!-\!Q(s^i, a^i|\omega^Q)\big)^2.
\end{aligned}
\end{equation}

The actor network can be trained with the policy gradient, which is described as follows:
\begin{equation}\label{policy_gra}
\begin{aligned}
	\triangledown_{\omega^\pi}J \approx \frac{1}{V} \sum\nolimits_{i=1}^{V} \triangledown_{a^{i}}Q(s^i, a^i|\omega^Q) \triangledown_{\omega^\pi} \pi(a^i,s^i|\omega^\pi).
\end{aligned}
\end{equation}

Finally, we update two target networks with rate $\eta$.

\section{Simulation Results}\label{simlresult}

In this section, simulation results are provided to evaluate the performance of our proposed algorithm. The BS and the RIS are located at (0, 0, 30 $\rm{m}$) and (100 $\rm{m}$, 20 $\rm{m}$, 10 $\rm{m}$), respectively. The users are  uniformly and randomly placed in a circle centered at (150 $\rm{m}$, 0, 1.5 $\rm{m}$) with radius of 20 $\rm{m}$. The large-scale path loss is given by $PL = P{L_0} - 10\alpha {\log _{10}}\left( {{d \mathord{\left/
 {\vphantom {d {{d_0}}}} \right.
 \kern-\nulldelimiterspace} {{d_0}}}} \right)$, where $P{L_0}$ is the path loss with the reference distance $d_0$, $\alpha$ is the path loss exponent, and $d$ is the communication distance. Unless otherwise stated, the simulation parameters are set as follows: Noise power density of -174 dBm/Hz, channel bandwidth of 1 MHz, reference path loss of $P{L_0}=-30\  {\rm{dB}}$, reference distance of $d_0=1$ $m$,  the number of antennas at the BS of $M=8$, Rician factors of $\delta=2.2$, $\varepsilon_k=3.75$ and $\eta_k=2.2$. There are a total of  $T=150$ CCTIs, and   $\tau_c=150$ time slots in each CCTI.

\begin{figure*}
	\centering
	\begin{minipage}[t]{0.33\linewidth}
		\centering
		\includegraphics[width=2.5in]{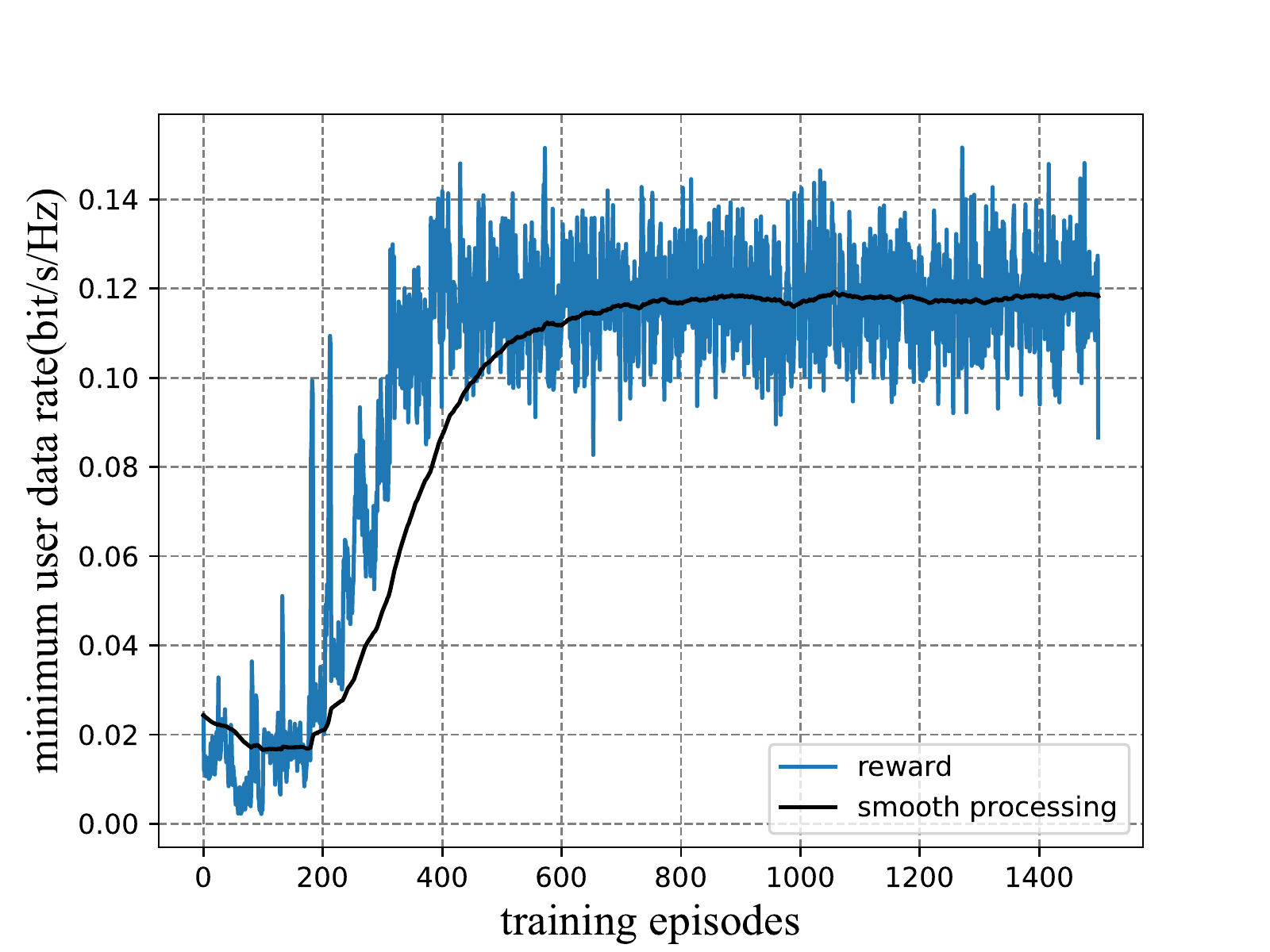}
		\caption{Convergence behaviour of the DDPG algorithm. \qquad}
		\label{convergence}
	\end{minipage}%
	\begin{minipage}[t]{0.33\linewidth}
		\centering
		\includegraphics[width=2.5in]{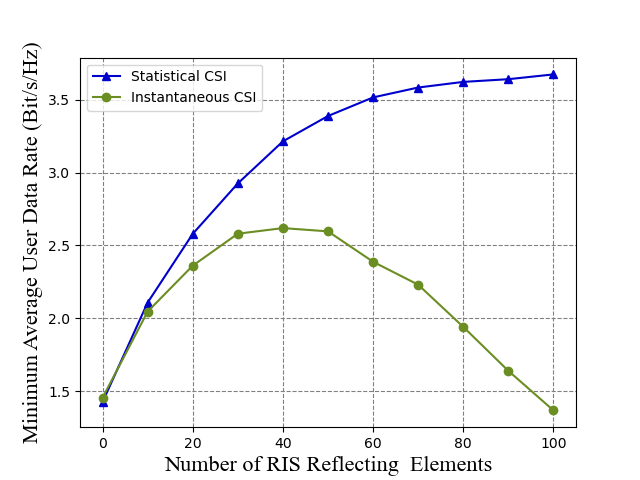}
		\caption{Minimum average user data rate versus the number of reflecting elements.}
		\label{compare_Rate}
	\end{minipage}
	\begin{minipage}[t]{0.33\linewidth}
		\centering
		\includegraphics[width=2.5in]{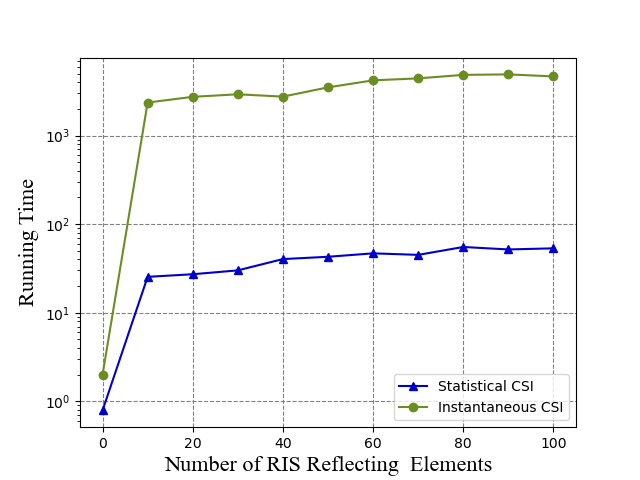}
		\caption{Minimum user data rate versus the number of reflecting elements.}
		\label{compare_Complexity}
	\end{minipage}
	%\vspace{-10pt}
\end{figure*}

In Fig.~\ref{convergence}, we first illustrate the convergence behaviour of the proposed DDPG algorithm. The numbers of reflecting elements and users  are  set as $N=80$ and $K=10$, respectively. The number of links from BS to RIS is $I=2$.  There are two curves in Fig.~\ref{convergence}. One curve named as `Reward'  is obtained by:
\begin{equation}\label{hiedhw}
  {\tilde r^{(t)}} = \mathop {\min }\limits_k \frac{1}{t}\sum\nolimits_{l = 1}^t {R_k^{(t)}}.
\end{equation}
Note that ${\tilde r^{(t)}}$ is different from that in (\ref{reward})  due to the fact that the latter has lower computational complexity in the training process. The second curve named as `Smooth processing' can be obtained as follows:
\begin{equation}\label{kjpyoj}
  r_{{\rm{smooth}}}^{(t)} = w*r_{{\rm{smooth}}}^{(t - 1)} + (1 - w)*{r^{(t)}}.
\end{equation} It can be observed from Fig.~\ref{convergence} that the DDPG converges rapidly and generally 1000 episodes are enough for the convergence.

In Fig.~\ref{compare_Rate}, we illustrate the minimum average user date (MAUR) versus the number of RIS reflecting elements. Four users are considered, and $I$ is set to one. Our proposed long-term CSI-based scheme is compared with the existing scheme that is based on instantaneous CSI \cite{Chongwen2020,keming2020}. The data rates for both schemes are calculated based on (\ref{rgsr}) and (\ref{rsfre}). Results in Fig.~\ref{compare_Rate} show that the MAUR achieved by the long-term CSI-based algorithm always increases with the number of reflecting elements. However, the MAUR achieved by the conventional instantaneous CSI-based scheme first increases with the number of the reflecting elements, and then decreases with it. The main reason is that, when the number of reflecting elements is small, the increased passive beamforming gain brought by the RIS outweigh the penalty due to the increased pilot overhead. However, when additionally increasing the number of reflecting elements, the detrimental effect of increased pilot overhead starts to dominate the system performance.

Finally, we provide Fig.~\ref{compare_Complexity}  to compare the computational complexity of the two algorithms.  It is observed that the total computational complexity of the proposed algorithm is much lower than the conventional instantaneous CSI-based algorithm. The reason is as follows. For the instantaneous CSI-based design, the algorithm needs to be implemented in each CCTI, while the statistical-CSI-based scheme only needs to  run the DDPG algorithm once in the considered time intervals.

\section{Conclusions}\label{conclu}

In this paper, we have proposed a DDPG algorithm to solve the RIS-aided multiuser communication systems, where both the beamforming vectors at the BS and the phase shifts at the RIS are designed based on the statistical CSI. Simulation results show that the proposed algorithm converges rapidly. In addition, compared with  the conventional instantaneous CSI-based scheme, the proposed algorithm has much lower computational complexity and higher net throughput.

\bibliographystyle{IEEEtran}

\bibliography{myre}

% that's all folks

\end{document}